\documentstyle[]{article}

\begin{document}
\begin{center}
{\Large \bf Extension of fixed point clustering: \\A cluster criterion}
\end{center}
\begin{center}
{\bf A. Hutt\footnote{e-mail:hutt@cns.mpg.de}, F. Kruggel\\}
Max Planck-Institute of Cognitive Neuroscience,
\\ Stephanstrasse 1a, 04103 Leipzig, Germany\\
\end{center}
The present report extends the method of fixed point clustering~\cite{PRERC} by introducing an indirect criterion for the number of clusters. The derived probability function allows an objective distinction of clustered data and data in between clusters. Applications on simulated data illustrate the clustering method and the probability function. 

\section{Introduction}
The dynamics of spatially extended systems can be measured by sets of multi-detector arrays. Most spatio-temporal analysis methods fitting multi-dimensional dynamical models \cite{Kirby,uhl-PRE,physicad-uhl,Ramsay,siegert,PRE} consider data over the full time range. In \cite{PRERC}, a method was described for partitioning spatio-temporal signals into time segments, in which the signal can be modeled by deterministic ordinary differential equations near fixed points. Each dynamical system is determined by a nonlinear spatio-temporal analysis \cite{PRE}. The earlier proposed algorithm in~\cite{PRERC} works with an arbitrary number of clusters $k$. Since results are dependant on $k$, an objective criterion for the number of clusters is necessary. The present report introduces a different segmentation algorithm and aims to derive an indirect criterion for the number of temporal segments. \\
In the present report, we use the K-Means algorithm~\cite{cluster-buch,Moody} for segmenting data, which addresses each data point to a single cluster. Since K-Means works with an arbitrary number of clusters and this number is crucial to clustering results, we derive a  probability function representing the degree of membership of a data point at time $t$ to a cluster. It addresses data to clusters or transition parts between clusters and hence determines the number of necessary clusters. Applications to simulated non-stationary data illustrates the probability measure.

\section{Fixed Point Clustering (FPC)}
In the following, a signal trajectory is assumed as compound of a sequence of segments governed by saddle point dynamics. Under the hypothesis, that these segments comprise the main functionality of the underlying system, we aim to extract them from the signal. Trajectories approach saddle points along
 their stable manifolds whereas they leave the vicinity of the fixed points
 along the unstable manifolds. The signal points accumulate close to the fixed
 points if the signal is sampled at a constant rate. This accumulation may also be regarded as a point cluster in data space. Subsequently, stable manifolds in multi-dimensional signals lead to point clusters and their detection can be treated as a recognition problem in data space \cite{PRERC}.

\subsection*{The clustering algorithm}
A $N$-dimensional spatio-temporal signal can be described by a data vector 
${\bf q}(t)\in\Re^N$, where the component $q_j(t_i)$ represents a data point
 at time $i$ and detection channel $j$. The clustering algorithm aims at 
cluster centers \{${\bf k}_k$\}, whose mean Euclidean distance to a set of 
 data points ${\bf q}(t_i)$ is minimal. The presented implementation follows
Moody et~al.\cite{Moody} and is sketched in Fig.~\ref{Abb2}.\\
Cluster centers ${\bf k}^0$ are initialized at random locations in the data and their Euclidean distances to each data point are calculated. K-Means defines memberships of data points to a cluster by the smallest Euclidean distance to its center. Thus, data are segmented into $k$ clusters and new cluster centers ${\bf k}^1$ are calculated as means of clustered data points. Distances between data points and centers ${\bf k}^n$ are re-estimated until a convergence condition is fulfilled. This criterion can be set either as a upper Euclidean distance limit between sequential cluster centers ${\bf k}^n, {\bf k}^{n+1}$ or as number of iterations. We choose to limit the number of iterations to $25$. 
%
%

\subsection*{Simulated spatio-temporal data and results}
Now, a low-dimensional simulated signal ${\bf A}(t)$ is introduced describing amplitudes of multi-dimensional spatial patterns ${\bf v}_i$ by
\begin{eqnarray*}
{\bf q}(t)=\sum_iA_i(t){\bf v}_i.
\end{eqnarray*}
This superposition describes a spatio-temporal signal ${\bf q}(t)$.\\
The dataset ${\bf A}(t)$ is generated by
\begin{eqnarray}
\dot{A}_1=\epsilon A_1-A_1[A_1^2+(2+b)A_2^2+(2-b)A_3^2]+\Gamma(t)\nonumber\\
\dot{A}_2=\epsilon A_2-A_2[A_2^2+(2+b)A_3^2+(2-b)A_1^1]+\Gamma(t)\label{kl}\\
\dot{A}_3=\epsilon A_3-A_3[A_3^2+(2+b)A_1^2+(2-b)A_2^2]+\Gamma(t)\nonumber.
\end{eqnarray}
Parameters are set to $\epsilon=1$, $b=2$ and $\Gamma(t)\in[-0.05...0.05]$ represents additive noise following a uniform deviate. Equations~\ref{kl} describe the convection onset of a Rayleigh-Benard-experiment in the presence of rotation\cite{PRERC,kli,Busse}.\\ 
A 3-dimensional trajectory ${\bf A}(t)$ is calculated by $2200$ integration steps with the 
initial condition ${\bf A}(t=0)=(0.03,0.2,0.8)$, see Fig.~\ref{Abb4}.
%
%
The trajectory passes the saddle points ${\bf A}_3^0=(0,0,1), {\bf A}_1^0=(1,0,0)$ and ${\bf A}_2^0=(0,1,0)$ in this sequence, and then returns to ${\bf A}_3^0$.\\
The K-Means algorithm is applied on the simulated data for different number of 
clusters $k=2,..,7$. In Fig.~\ref{Abb7}, the Euclidean distances from each 
data point to the determined cluster centers are plotted in  
temporal sequence for each $k$. When a trajectory approaches or moves from 
a cluster center, its Euclidean distance to the center decreases resp. 
increases. These changes can be observed in Fig.~\ref{Abb7}. For fixed number
 of clusters, each data point is considered to be member of a cluster, 
whose center is closest to the data point. \\
Comparing obtained clustering results for different $k$, clustered time windows [0;$\sim$350], [$\sim$350;$\sim$1050], [$\sim$1160;$\sim$1610] and [$\sim$1740;2200] are recognized, which borders remain similar for different $k$.
%
%

\section{The cluster criterion}
Although there might be only a limited number of clusters $k_d<k$ in the data,
 K-Means determines $k$ clusters also including void clusters. In Fig.~\ref{Abb7}, small clustered time windows are visible, whose occurences and temporal widths strongly depend on $k$. They are considered as invalid clusters. Conversely, a first qualitative criterion for valid clusters may be formulated as: 
\begin{itemize}
\item cluster widths and locations in time remain independent of $k$ and
\item the Euclidean distances of clustered data points to centers is 
obviously smaller then the Euclidean distances of points to the next nearest 
cluster center and 
\item the width of the clustered time window is not too small.
\end{itemize}

Although these criteria are rather heuristic than formal, they proved to be 
useful in practice~\cite{PRERC}.
Now, we try to evolve them quantitatively. The first item can be formulated 
as a sum over all clustering results: valid contributions are additive if they 
occur for all $k$, others vanish in the sum as small contributions. Thus the
 contribution of a valid cluster to the sum should be large, not reliable 
clusters should contribute with small values. A good quantity for these 
contributions is the area between the curves of  the signal-nearest
 cluster-distance and signal-next cluster-distance. This definition allows 
the analytical formulation of the second item and is outlined in 
Fig.~\ref{Abb8}. Each data point $t_i$ obtains an index corresponding to the 
cluster $j$ it is member of. The index is equal the relativ area 
$\frac{A_j^{(k)}(t_i)}{T\sum_jA_j^{(k)}}$, where $T$ denotes the number of data
 points. By summing up the indices over $K$ cluster realizations for every
 data point, a degree of membership $M(t)$ for every data point is obtained:
\begin{eqnarray*}
A_r^{(k)}(t_i)&=&\frac{A_j^{(k)}(t_i)}{\sum_jA_j^{(k)}}\\
M(t_i)&=&\frac{\sum_{k=2}^{K\le T}A_r^{(k)}(t_i)}{\sum_{i=1}^T\sum_{k=2}^{K\le T}A_r^{(k)}(t_i)} .
\end{eqnarray*}
$M(t)$ represents a probability, that a data point at time $t$ belongs to a cluster. 
The application to simulated data with $K=30$ leads to results of $M(t)$ shown in Fig.~\ref{Abb9}. Clustered time windows can be recognized as regions of high values of $M(t)$. Four plateaus of $M(t)$ are recognized at [0;250], [340;1010], [1180;1560] and [1750;2200] with borders at drastic value changes. Regions between these plateaus are considered as non-functional transitions parts. Comparing time windows in the original signal(Fig.~\ref{Abb4}) and detected clustered time windows in Fig.~\ref{Abb7} and Fig.~\ref{Abb9}, good accordance of time windows near fixed points and cluster results are recognized.\\
%
%

%
%

\section{Conclusion}
The present brief report extends the fixed point clustering method by introducing a probability function $M(t)$. High values of $M(t)$ indicate clustered data points. Fixed point clustering relates temporal dynamics near fixed points showing attractive and repelling properties with clusters in dataspace. By the presented extension, regions in data space near such fixed points can be determined independant of the number of clusters. Applications to spatio-temporal signals in hydrodynamics, metereology or brain science~\cite{Psycho,dpg} are possible.

\newpage
\begin{figure}[h]
\begin{center}
\caption{The implementation steps of the K-Means algorithm.\label{Abb2}}
\end{center}
\end{figure}

\begin{figure}[p]
\begin{center}
\end{center} 
\caption{Trajectory of the 3-dimensional signal ${\bf A}(t)$. It starts near a saddle point and passes two others, before it returns to the initial saddle point. The numbers denote the timesteps of the trajectory at their locations. \label{Abb4}}
\end{figure}

\begin{figure}[h]
\begin{center}
\caption{Cluster results for $k=2,..,7$. The Euclidean distances between data points and detected clusters are shown.\label{Abb7}}
\end{center}
\end{figure}

\begin{figure}[h]
\begin{center}
\caption{Sketch to illustrate the introduced criterion of a clusters validity. Area $A_j$ between two distance curves indexes the data points, which belong to cluster $j$. Large areas indicates at a high degree of membership.\label{Abb8}}
\end{center}
\end{figure}

\begin{figure}[h]
\begin{center}
\caption{Degree of membership $M(t)$ for every data point as a sum of $K=30$ clustering results. Plateaus denote valid clusters, which are delimited by rapid changes.\label{Abb9}}
\end{center}
\end{figure}

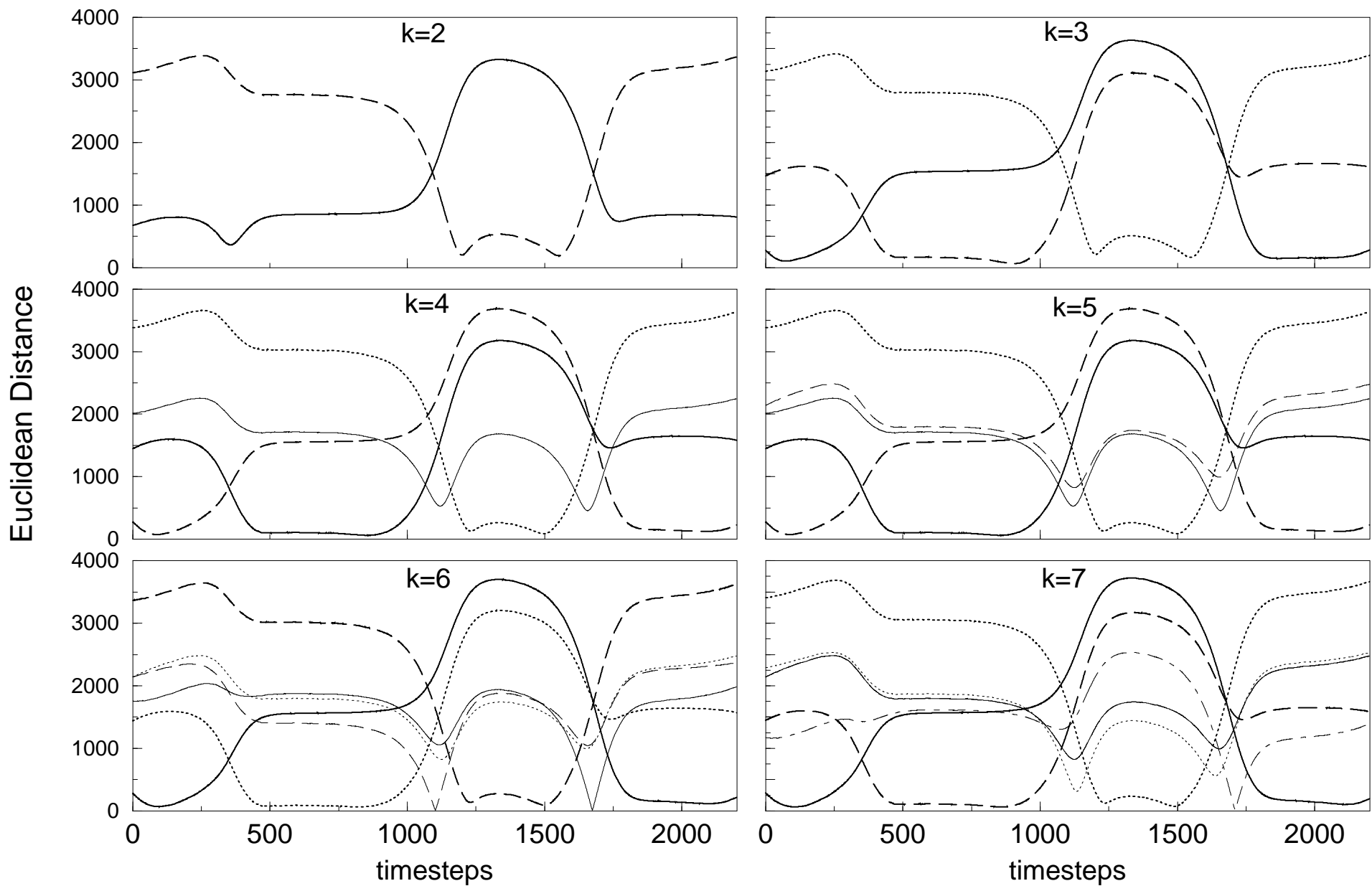
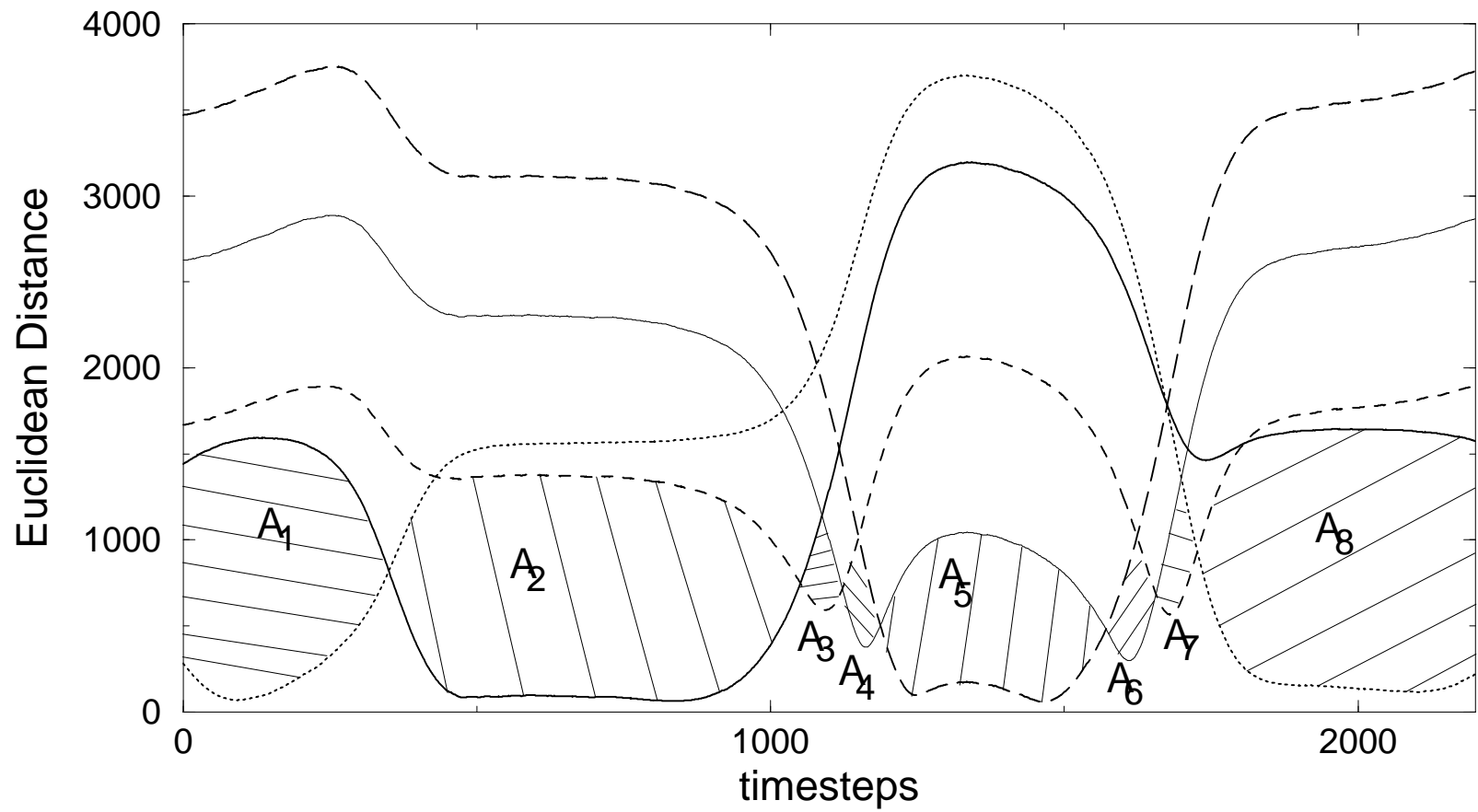
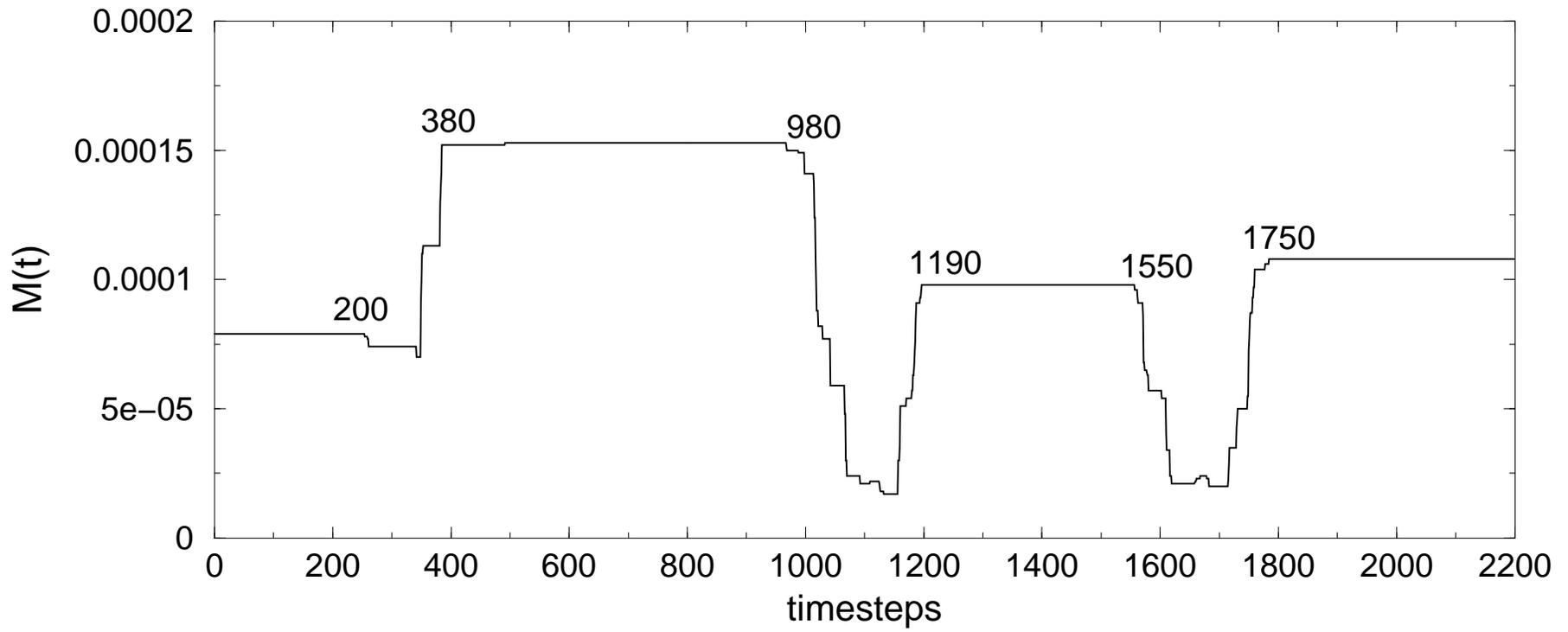
\end{document}